\documentclass[10pt,aps,prb,twocolumn,amsmath,amssymb,floatfix]{revtex4-2}

\usepackage[dvipsnames]{xcolor}
\usepackage{graphicx}
\usepackage[
colorlinks,
citecolor=blue,
linkcolor=blue,
urlcolor=blue,plainpages=false,pdfpagelabels
]{hyperref}
\usepackage{txfonts}
\usepackage{amsmath,bm}

\newcommand{\bra}[1]{\ensuremath{\langle{#1}|\,}}
\newcommand{\ket}[1]{\ensuremath{\,|{#1}\rangle}}
\newcommand{\braket}[1]{\ensuremath{\langle{#1}\rangle}}

\newcommand{\beq}{\begin{equation}}
\newcommand{\eeq}{\end{equation}}
\newcommand{\bea}{\begin{eqnarray}}
\newcommand{\eea}{\end{eqnarray}}

\begin{document}
	\title{Quantum friction in the Hydrodynamic Model}
	
	\author{Kunmin Wu}
	\affiliation{Department of Physics
		and Materials Science, University of Luxembourg, 1511 Luxembourg, Luxembourg}
	\author{Thomas L. Schmidt}
	\affiliation{Department of Physics
		and Materials Science, University of Luxembourg, 1511 Luxembourg, Luxembourg}
	\author{M. Bel\'en Farias}
	\affiliation{Department of Physics
		and Materials Science, University of Luxembourg, 1511 Luxembourg, Luxembourg}
	
	\date{\today}
	
	\begin{abstract}
		In this work, we study the phenomenon of quantum friction in a system consisting on an atom
		moving at a constant speed parallel to a metallic plate.
		We use a hydrodynamic model to describe the
		degrees of freedom of a clean metal without internal dissipation. We model the polarizable atom as a two-level system
		with a unique ($l=0$) ground state and a three-fold degenerate ($l=1$) excited state. We show that a quantum frictional force is present even in the absence of
		intrinsic damping in the metal, but that there is a threshold in the relative velocity that
		give rise to such a force. In particular, we find that for friction to occur, the atom must move at a velocity
		larger than the effective speed of sound in the material, a condition which can be reached near empty or filled bands. We provide analytical arguments to show that this result holds
		at all orders in perturbation theory.		
	\end{abstract}
	
	\maketitle
	
	\section{Introduction} \label{sec:int}

	Quantum friction, also known as Casimir friction or van der Waals friction, was one of the most debated consequences of the fluctuations of the electromagnetic (EM) vacuum field \cite{pendry_debate,philbin2009no}.
	This phenomenon is a non-equilibrium counterpart of the static Casimir effect \cite{casimir1948,milton2004casimir},
	and describes a force against the direction of motion if two neutral objects move relative to each other at a finite speed. The standard setup for
	the study of this force consists of two objects moving parallel to each other at a constant velocity \cite{milton2016reviewfriction, barton_atom_halfspace,pendry97,volokitin_persson}, but it was also studied in cases with acceleration \cite{intravaia_acceleration,farias2019motion} and non-parallel motion \cite{volokitin2003resonant, klatt_farias}. In contrast to the dynamical Casimir effect \cite{review_dyncas}, no acceleration is needed to produce the force, and no real photons are excited out of the vacuum. Instead, the origin of the quantum friction force can be traced back to virtual Doppler-shifted photons exchanged between the two objects, thus exciting their internal degrees of freedom and leading to dissipation.
	
	This force, however, is extremely small and short-ranged, and has thus been eluding experimental detection.
	To address this problem, many theoretical works were dedicated to its study in the past years. Research has mainly progressed in two different directions.
	On the one hand, different theoretical methods have been explored for the study of this effect. Since the system is in a non-equilibrium state, methods
	that work reliably with common fluctuation problems cannot be applied to quantum friction. Among the theoretical approaches
	that have been employed, we can name functional methods based on path integrals \cite{farias_friction}, time-dependent perturbation theory \cite{barton_atom_halfspace}, macroscopic quantum electrodynamics \cite{klatt2016} and generalized fluctuation theorems \cite{dalvit_intravaia,reiche2020nonequilibrium}.
	
	On the other hand, different systems have been proposed with the goal of obtaining an enhancement of the frictional force. The interplay
	between quantum friction and decoherence \cite{farias2016,viotti2019thermal} and its influence on the Berry phase \cite{farias2020towards} have been explored
	as a way of indirectly detecting traces of the phenomenon without actually measuring the force. Different materials, such as graphene \cite{volokitin2011quantum,farias_graphene,shaukat2020drag} and topological materials \cite{farias2018quantum}, which have already been shown to enhance vacuum fluctuation forces like Casimir forces \cite{wilson2015repulsive,farias2020}, have been proposed as possible platforms in which the quantum friction force
	might be enhanced.
	
	In this paper, we will approach the problem from a different angle. We shall consider a system consisting on an atom moving at constant speed above a planar metallic slab. We will study the interaction between the atom's electric dipole moment and the fluctuating electromagnetic field, which is in turn modified by the presence of the material.
	The metal will be modelled from a microscopic point of view
	using the hydrodynamic (HD) model, following the review by G.~Barton \cite{barton1979hydrodynamicmodel}. The HD model has the advantage of
	accounting exactly for the non-local Coulomb interactions inside the metal without excessive complexity. Indeed, the spatial dispersion becomes relevant at small distances from the surface and when the charge carriers in the material cover distances larger than the interatomic separation, and it has been proven particularly relevant for quantum friction \cite{reiche2017spatial}. Another characteristic of the HD model is that it does not include an explicit phenomenological damping, unlike for instance the impurity scattering which is the basis of the Drude model. This makes it possible to study the mechanisms giving rise to friction
	even if the metal itself has no intrinsic dissipation, as was seen e.g.~in graphene \cite{farias_graphene}.
	
	In this work, we will combine the
	HD model for the material with an approach for quantum friction based on a dipole interaction with the atom and time-dependent perturbation theory \cite{barton_atom_halfspace}. This approach allows a relatively simple analytical treatment, which starts from the equations of motion of the fields
	and the microscopic degrees of freedom of the material, and will shed some light onto the microscopic mechanisms behind this contactless dissipation.

	The paper is organized as follows. In Sec.~\ref{sec:HDmodel} we will briefly review the HD model and obtain expressions for the ``dressed'' electromagnetic field in the vicinity of a semi-infinite metallic plate, following the calculations in Ref.~\cite{barton1979hydrodynamicmodel}.
	In Sec.~\ref{sec:system} we will introduce the system under study and the Hamiltonian of the composite system, which is formed by the particle, the plate and the EM field. In Sec.~\ref{sec:state}, we will use time-dependent perturbation theory to find
	the state of the system as a function of time. We will then find the frictional force from this quantum state in Secs.~\ref{sec:2ndorder} and \ref{sec:4thorder}. Lastly, we will present our conclusions in Sec.~\ref{sec:conc}.

	\begin{figure}[t]
		\includegraphics[width=\columnwidth]{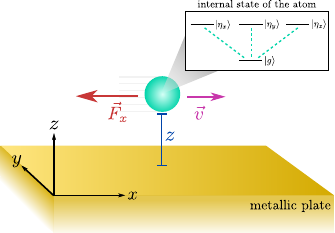}
		\caption{\label{fig:system} (color online): Schematic representation of the system under study. An atom at a distance $z$ over a metallic plate moves at constant speed $v$ parallel to the surface. The quantum frictional force opposes the motion of the atom. \textit{Inset:} the internal state of the atom is a two-level system for which the excited state is threefold degenerate, in analogy with the $1s$ and $2p$ levels of the hydrogen atom.}
	\end{figure}

	\section{The hydrodynamic model} \label{sec:HDmodel}
	
	\subsection{Differential equations of the model}
	
	In the HD model, electrons in a material are modeled in analogy with a fluid, retaining the long-range interactions of an electron gas, but not the particle aspects.
	In a material with electron number density $n$ and electron mass $m$, a displacement field $\bm{\xi}$ of the electrons from their equilibrium positions
	will result in a deviation of the density $\Delta n = -n \boldsymbol{\nabla \cdot \xi}$ as well as in a pressure $\Delta P$.
	If we consider elastic collisions between electrons, the pressure will obey the adiabatic polytropic equation
	$PV^\Gamma = \text{const.}$, where $\Gamma$ is the ratio of specific heats. This polytropic equation can be
	cast in the form $P \propto \rho^\Gamma = (m n)^\Gamma$, which allows us to write
	\begin{align}
	\Delta P &= \frac{dP(n)}{dn} \Delta n = \left(\Gamma \frac{P}{n}\right) \Delta n= m\beta^2 \Delta n \nonumber \\
	&= -nm \beta ^2  \boldsymbol{\nabla \cdot \xi} \, . \label{eq:how_to_deduce_pressure}
	\end{align}
	Here, we have identified the speed of sound or compressional wave speed $\beta$, defined for classical electrohydrodynamical waves
	as $\beta^2 = \Gamma P / (m n)$ \cite{book_jackson}.
	It can be understood as how fast the electrons can affect their neighbors, and its magnitude is of the order of the Fermi velocity {\cite{barton_atom_halfspace}.
	
	A deviation $\Delta n$ in the electron density will generate an electric field, which in the non-retarded limit (where the speed of light $c \rightarrow \infty$), is given by Gauss's law,
	\begin{align}  \label{eq:Maxwell_eq}
	\boldsymbol{ \nabla \cdot E} &= 4 \pi e \Delta n =  -4 \pi n e \; \boldsymbol{\nabla \cdot \xi} \, .
	\end{align}
	In general, the displacement vector filed $\boldsymbol{\xi}$ can be decomposed into an irrotational part which can be expressed as the gradient of a scalar potential, $\boldsymbol{ \nabla } \Psi$, as well as a rotational part $\boldsymbol{ \nabla } \times \boldsymbol{A}$. However, a rotational contribution in $\boldsymbol{\xi}$ does not affect the Coulomb interactions, so its effect can be neglected \cite{barton1979hydrodynamicmodel}.
	As a consequence, we can use ${\boldsymbol{\xi} = - \boldsymbol{\nabla} \Psi}$. Hence, the previous equation can be rewritten as follows in terms of the conventional electrostatic potential $\Phi$, related to the electric field via $\boldsymbol{E} = \boldsymbol{ \nabla } \Phi$, and the displacement potential $\Psi$,
	\begin{equation}
	\nabla^2 \Phi = -4 \pi ne \nabla^2 \Psi. \label{eq:relation_of_potentials}
	\end{equation}
	Furthermore, we use the equation of motion for the electronic displacement field ${n m \boldsymbol{ \ddot{\xi}}= n e \boldsymbol{E} - \boldsymbol{\nabla} (\Delta P)}$. Combining it with the previously introduced potentials and using an oscillatory time-dependence of the form $\boldsymbol{\xi}( t) \propto e^{- i \Omega t}$, corresponding to a single mode with frequency $\Omega$, we obtain
	\begin{align}
	\Phi &= - \frac{m}{e} (\Omega^2 + \beta^2 \nabla^2) \Psi. \label{eq:transformed_eq_of_motion}
	\end{align}
	Combining this equation with Gauss's law, we arrive at a differential equation for the displacement potential
	\begin{equation}
	\nabla^2 (\Omega^2 - \omega_p^2 + \beta^2 \nabla^2) \Psi = 0, \label{eq:original_DE}
	\end{equation}
	where $\omega_p = 4 \pi n e^2/m$ is the plasma frequency.

	We are interested in the study of a system in which a metal described by the HD model
	fills the half-space $z < 0$ (see Fig.~\ref{fig:system}) while vacuum fills the rest of the space. Inside the material ($z<0$), the behaviour of the fields is governed by the equations introduced in the previous paragraphs, in particular Eq.~\eqref{eq:original_DE}. Outside it ($z>0$), there are no free charges, so the Laplace equation dictates that $\nabla ^2 \Phi =0$.
	
	At the interface ($z=0$), we need to match the solutions of these equations using the proper boundary conditions. The electrostatic potential and the normal component of the electric field must be continuous at the interface, resulting in $\Phi|_{z=0_-} = \Phi|_{z=0_+}$ and $\partial_z \Phi |_{z=0_-} = \partial_z \Phi |_{z=0_+}$. For the displacement field, we impose the condition that
	no electron can escape the plate, which means that $\xi_z|_{z=0_-} = -\partial_z \Psi|_{z=0_-} = 0$.
	
	The vacuum modes of the EM field do not cause quantum friction. Rather, it is dominated by the near-field effects at the interface. Therefore, we can concentrate on evanescent waves which vanish at large distances from the interface, and therefore demand that $\Phi(z \rightarrow \pm \infty) =
	\Psi(z \rightarrow - \infty) = 0$.
	
	Moreover, due to the rotational symmetry in the $x-y$ plane, we can make the ansatz $\Psi(z\le 0) = e^{i\textbf{k}\cdot\bm{\rho}} \psi(z)$ and $\Phi = e^{i{\textbf{k}\cdot\bm{\rho}}} \phi(z)$, with $\bm{\rho}=(x,y)$ and $\textbf{k}=(k_x,k_y)$.
	With this we can rewrite Eq.~\eqref{eq:original_DE} for the fields inside the plate as
	\begin{align}
	&\left(-k^2+\frac{d^2\;}{dz^2}\right)\left(\Omega^2-\omega_p^2-\beta^2k^2+\beta^2\frac{d^2\;}{dz^2}\right)\psi(z)= 0 \, ,  \label{eq:main_DE}
	\end{align}
	where $k^2 =k_x^2+k_y^2$.
	On the other hand, in the vacuum outside the material, the normalizable solutions of the Laplace equation are evanescent waves and have the form
	\begin{align}
	&\phi(z) = \phi(z=0_+)\; e^{-kz} \;\;\;\; (z\geq 0) \, .  \label{eq:vacuum_solution}
	\end{align}
	where the constant term $\phi(z=0_+)$ is fixed by the continuity of potentials at the interface, once the solution inside the material has been determined.
	
	\subsection{Surface modes and bulk modes}

	To solve Eq.~\eqref{eq:main_DE}, we make an exponential ansatz proportional to $e^{p z}$ with $\text{Re } p\geq 0$ to ensure convergence inside the material. This leads to non-trivial solutions that must
	fulfill
	\begin{align}
	p^2 &= \frac{1}{\beta^2} (-\Omega^2 +\omega_p^2+\beta^2k^2) \label{eq:non_local_solution_p} \, .
	\end{align}
	The solutions with $p$ determined by Eq.~\eqref{eq:non_local_solution_p}
	give rise to two different types of modes: surface modes and bulk modes.
	The \textit{surface modes} exist when $\Omega ^2<\omega_p^2+\beta^2k^2$, because this results
	in a real-valued $p$ and thus a decaying function inside the material, while \textit{bulk modes} are present when $\Omega ^2 \geq \omega_p^2+\beta^2k^2$ such that $p$ becomes purely imaginary. Physically, we can thus think of $\sqrt{\omega_p^2+\beta^2k^2}$ as an effective plasma frequency in the HD model.
	We define the frequencies $\Omega_s$ and $\Omega_b$ of the surface modes and the bulk modes, respectively, as
	\begin{align}
	\Omega_s^2 &\equiv \omega_p^2+\beta^2k^2-\beta^2p_s^2, \label{eq:surface_modes_condition}\\
	\Omega_b^2 &\equiv \omega_p^2+\beta^2k^2+\beta^2p_b^2 \label{eq:bulk_modes_condition}.
	\end{align}
	with real $p_s$ and $p_b$. After imposing the boundary conditions, the respective solutions for surface and bulk modes have the form
	\begin{align}
	\psi_s(z\leq 0) &= N_k (p_se^{kz}-ke^{p_sz}), \label{eq: surface solution1}\\
	\psi_b(z\leq 0) &= M_k \left\{\cos(p_bz)+ C\left[e^{kz}-\frac{k}{p_b}\sin(p_bz)\right]\right\}, \label{eq:bulk solution1}
	\end{align}
	where $N_k$ and $M_k$ are normalization constants and $C$ is another constant to be determined.
	
	Since our goal is to obtain the electric potential, we can insert the solution $\psi_s$ for surface modes into Eq.~\eqref{eq:relation_of_potentials} to find
	\begin{equation}
	\phi_s(z\leq 0) =-\frac{m}{e}N_k(\Omega_s^2p_se^{kz}-\omega_p^2ke^{p_sz}). \label{eq: surface solution2}
	\end{equation}
	By imposing the boundary conditions, we arrive at
	\begin{align}
	\phi_s (z > 0) &= -\frac{m}{e}N_k(\Omega_s^2p_s-\omega_p^2k)e^{-kz}, \label{eq: vacuum surface solution} \\
    p_s &= \frac{1}{2}\left(-k +\sqrt{k^2+\frac{2\omega_p^2}{\beta^2}}\right) \label{eq:ps solution}.
	\end{align}
	Therefore, the dispersion relation for surface modes reads
	\begin{equation}
	\Omega_s(k) = \frac{1}{2}\left(\sqrt{2\omega_p^2+\beta^2k^2} + \beta k\right). \label{eq:dispersion_relation}
	\end{equation}
	With these results we have determined the classical electrostatic potential generated by the surface modes outside the material up to a normalization
	constant. Since quantum friction is a low-frequency phenomenon, these modes will play an important role in the upcoming calculations.
	
	For the bulk modes, the electrostatic potential outside the plate becomes, after applying the boundary conditions,
	\begin{align}
	\phi_b(z >0) &= -\frac{m}{e}M_k\omega_p^2\left(\frac{\Omega_b^2-\omega_p^2}{2\Omega_b^2-\omega_p^2}\right)e^{-kz} \label{eq: vaccum bulk solution2}.
	\end{align}
	This is the electrostatic potential generated by bulk modes with frequencies larger than $\sqrt{\omega_p^2+\beta^2k^2}$. Every bulk mode will be determined by the two independent variables $k$ and $p$. The classical solution for the electrostatic potential field outside the material would be given by the sum of Eqs. \eqref{eq: surface solution2} and \eqref{eq: vaccum bulk solution2}.

	\subsection{Quantization}
	Using the classical field outside the material, we derive the quantum field using canonical quantization. After normalizing, it becomes
    \begin{widetext}
	\begin{align}
	\hat{\Phi}(z \geq 0) &= - \int d^2 k e^{i\textbf{k}\cdot \bm{\rho}-kz} \hat{a}_{\textbf{k}} \omega_p^2\sqrt{\frac{1}{4\pi k\Omega_s(\omega_p^2+2\Omega_s^2)}}
	-\int d^2 k\int dp e^{i\textbf{k}\cdot \bm{\rho}-kz} \hat{a}_{\textbf{k}p} \frac{\beta^2p_b\omega_p}{\pi\sqrt{\Omega_b[(2\Omega_b^2-\omega_p^2)^2 - 4\beta^2k^2\Omega_b^2]}} + \text{h.c.} \label{eq: quantized phi in vacuum},
	\end{align}
    \end{widetext}
	where the creation and annihilation operators obey the usual bosonic commutation rules:
	\begin{align}
	\left[\hat{a}_{\textbf{k}}, \hat{a}_{\textbf{k}'}^{\dagger}\right] &= \delta(\boldsymbol{k'}-\textbf{k}), \\
	\left[\hat{a}_{\textbf{k}p}, \hat{a}_{\textbf{k}'p'}^{\dagger}\right] &= \delta(\boldsymbol{k'}-\textbf{k}) \delta(p'-p).
	\end{align}
	These operators create or annihilate a composite excitation of both the electron fluid and the EM field, which one can think of as either a dressed photon or a plasmon polariton. It is common in the literature to find the excitation of the field referred to plainly as photons, but it is important to remember that these are not free photons. These are joint excitations of both the vacuum EM field and the internal degrees of freedom of the material.
	
	Since quantum friction is a low-frequency phenomenon (i.e., the contribution to the lowest order in the velocity can be obtained from the zero-frequency response of both the material and the atom \cite{dalvit_intravaia}), and we are in the non-retarded and near-field regimes,
	small frequencies will have a dominant effect and hence only surface modes will be relevant to our work. In the interaction picture,
	the quantized vacuum field becomes
	\begin{align}
	\hat{\Phi}(\textbf{r}, t) = - \int d^2 k\ e^{i\textbf{k}\cdot \bm{\rho}-i\Omega_s(k)t-kz} \hat{a}_{\textbf{k}}(t)\phi_k + \text{h.c.},  \label{eq:field_operator}
	\end{align}
	where $\phi_k^2 = \omega_p^4/[4\pi k\Omega_s(\omega_p^2+2\Omega_s^2)]$ encodes the information regarding the internal degrees of freedom of the material, and has units of velocity.

	\section{Interaction with the atom} \label{sec:system}

	We study an atom that moves at constant speed $v$ in front of a metallic plate. We assume that during the time the atom interacts with the metal, its distance $z$ to the plate can be approximated as constant. A schematic picture of the setup is shown in Fig.~\ref{fig:system}. Without loss of generality, we will assume that the velocity of the atom
	is along the $\hat{x}$ direction. We will assume $z$ to be on the order of nanometers to microns and $v$ much smaller than the speed of light, to ensure that we remain in the near-field and non-retarded regimes.
	
    The unique ground state of the atom corresponds to an unpolarized state, say the $1s$ orbital of a hydrogen-like atom. The lowest excited state which can be reached by an optical transition, say the $2p$ orbital, is three-fold degenerate and has an energy $\omega_b$. These lowest energy levels can be modelled as the ground state and first excited state, respectively, of a three-dimensional harmonic oscillator \cite{barton_atom_halfspace}. The polarization of the atom is then determined by the linear combination of $2p$ states into which the system is excited, and the polarization state $\ket{\bm{\eta}}$ of the atom can be uniquely labelled by a normalized three-dimensional polarization vector $\bm{\eta}$. The unit vectors $\bm{\eta}$ have components that fulfil the identity $\sum_{\bm{\eta}} \eta_i \eta_j = \delta_{ij}$, where the sum is over the three vectors of a chosen orthonormal basis. This situation is depicted as well in the inset of Fig.~\ref{fig:system}.
	
	The atom will interact with the EM field through its dipole moment $\bm{\hat{D}}$. This operator has vanishing expectation values
	on all eigenstates of the atom, but presents non-vanishing transition amplitudes. In the interaction picture, they are given by
	\begin{align}
	\braket{g \;\vert\hat{\boldsymbol{D}}\vert\;\boldsymbol{\eta}} = \boldsymbol{\eta}\, d\, e^{-i\omega_bt} \label{eq: excite the electron in the particle}, \;\;\;\;
	\braket{\boldsymbol{\eta}\;\vert\hat{\boldsymbol{D}}\vert\;g} = \boldsymbol{\eta}\, d\, e^{+i\omega_bt} \, ,
	\end{align}
    where $d$ denotes the strength of the dipolar coupling. In the dipolar approximation, the effect of an external electric potential can be described by adding an interaction term to the (free) Hamiltonian of the atom,
	\begin{align}
	\hat{V}(t) &= -\hat{\boldsymbol{D}}\cdot \left(-\boldsymbol{\nabla}\hat{\Phi}(t)\right), \label{eq:Vt}
	\end{align}
	where $\hat{\Phi}$ is the quantized electric potential in the vacuum at the position of the atom, given by Eq.~\eqref{eq:field_operator}. The frictional force acting on the particle will then be given by
	\begin{align}
	F_x(t) = \bra{\psi(t)} \hat{F}_x(t) \ket{\psi(t)},
	\end{align}
	where $\ket{\psi(t)}$ is the state of the system at time $t$, and $\hat{F}_x(t) = -\frac{\partial}{\partial x}\hat{V}(t)$
	is the force operator in the direction of motion, which is given by
	\begin{align}
	\hat{F}_x(t) &= -\int_{0}^{\infty} dk \int_{0}^{2\pi} d\theta \; \left(\hat{\boldsymbol{D}}(t)\cdot \textbf{k}\right) k^2\cos(\theta) \nonumber \\
	& \times e^{i{\textbf{k} \cdot \bm{\rho}(t)}-i\Omega_s t-k z(t)}\hat{a}_{\textbf{k}}(t) \phi_k + \text{h.c.}\;, \label{eq:general force operator}
	\end{align}
	where $\textbf{k}= (k_x,k_y)=(k \cos \theta , k \sin \theta)$. In our case, the trajectory of the particle will be given by
	$\bm{\rho}(t) = v \, t \, \hat{x}$, $z(t) = z$, with $z$ a constant. Hence, we have
	\begin{align}
	\hat{F}_x(t) &= -\int_{0}^{\infty} dk \int_{0}^{2\pi} d\theta \; \left(\hat{\boldsymbol{D}}(t)\cdot \textbf{k}\right) k^2\cos(\theta)\nonumber \\
	\;\;\;\;\;\;\;\;\;\;& \times e^{ik\cos(\theta)vt-i\Omega_st-k a}\hat{a}_{k}(t) \phi_k +h.c. \label{eq:the force operator}
	\end{align}
	To calculate the expectation value of the friction force, we need to compute the state of the system at a given time $t$. We shall do so in the next Section.
	
	%
	%
	%
	\section{Time-evolution of the system state} \label{sec:state}
	%
	
	Our complete system consists on two parts, the atom and the field, which interact via the dipolar
	potential $\hat{V}(t)$ defined in Eq.~\eqref{eq:Vt}. This interaction potential will act as a time-dependent perturbation to the system.
	This allows us to write the state of the system at a time $t$ by means of time-dependent perturbation theory as
	\begin{align}
	\ket{\Psi(t)} &=  (1 + c^{(2)}_0(t))\ket{g, 0} \nonumber \\
	&+ \sum_{\eta}\int d^2 k \left(c_1^{(1)}(t) +c_1^{(3)}(t)\right)\ket{\boldsymbol{\eta}, \textbf{k}} \nonumber \\
	& + \frac{1}{2!}\int d \textbf{k}_1 d\textbf{k}_2\; c_2^{(2)}(t)\ket{g, \textbf{k}_1\textbf{k}_2}+ \ldots, \label{eq: Atom+Field States}
	\end{align}
	where we are using a basis for the atom-photon state in which the atom can either be in its ground or excited state, and which contains a certain number of (dressed) photons with momenta $\textbf{k}_1, \textbf{k}_2, ...$. In particular, the states appearing in the previous expression are: the atom
	in its ground state and no photons ($\ket{g,0}$), the atom in an excited state and one photon with momentum $\textbf{k}$ ($\ket{\bm{\eta},\textbf{k}}$),
	and the atom in the ground state plus two photons, one with momentum $\textbf{k}_1$ and one with $\textbf{k}_2$ ($\ket{g, \textbf{k}_1 \textbf{k}_2}$). It can be shown that all other coefficients up to order $d^3$ will vanish.
	Due to the dispersion relation \eqref{eq:dispersion_relation}, the state of the dressed photons is determined solely by their wavenumber.
	
	Next, we will present the perturbative coefficients in Eq.~\eqref{eq: Atom+Field States}, that we will use later to calculate the frictional force. The first one is given by
	\begin{align}
	c_1^{(1)}(t) &= \frac{id \left(\boldsymbol{\eta}\cdot \vec{k}\right)^*\left(\phi_k\right)^{*}e^{-kz}}{\omega_b+\omega'_k - i\lambda} e^{i(\omega_b + \omega'_k)t}, \label{eq:c_11}
	\end{align}
	where $\lambda$ is a positive infinitesimal constant which is added to make the time integral converge at $t=-\infty$, and
	\begin{equation}
	\omega'_{k} = \Omega_s(k)-k\, v \, \cos(\theta) \label{eq:Doppler_Shifted_frequency}
	\end{equation}
	is the Doppler-shifted frequency.
	
	The coefficient $c_0^{(2)}(t)$, which corrects the contribution to the ground state, can be approximated as \cite{dalvit_intravaia}
	\begin{align}
	c_0^{(2)}(t) &\approx\; id^2\sum_{\eta}\int d^2k \frac{ \left|\boldsymbol{\eta} \cdot \vec{k}\right|^2 \left|\phi_k\right|^2 e^{-2kz}  t }{\omega_b+\Omega_s(k)-k\cos(\theta)v - i\lambda} \label{eq:c_02_complicated_form} \\
	& \equiv \; -i\delta \omega_gt - \frac{\gamma_gt}{2}. \label{eq:c_02}
	\end{align}
	where $\delta \omega_g$ is the frequency shift and $\gamma_g$ is the decay rate of the ground state,
	which can be obtained from Fermi's golden rule.
	\begin{align} \label{eq:gamma}
	\gamma_g &= \sum_{\eta}\int d^2k \; \left| {V}_{01}(t)\right|^2 \delta(\omega_b+\omega'_k),
	\end{align}
	where ${V}_{01} =  \braket{\boldsymbol{\eta}, \textbf{k} \vert \hat{V} \vert g,0}$ is the
	transition amplitude from a state with the atom in its ground state and no dressed photons, to a state with the atom in its excited state and one photon with momentum \textbf{k}.
	
	For $c_2^{(2)}(t)$, which involves processes with two photons, we have
	\begin{align}
	c_2^{(2)}(t) &= - \frac{d^2\left(\vec{k}_2 \cdot \vec{k}_1 \right)^*\left(\phi_{k_1}\phi_{k_2}\right)^*e^{-k_1z-k_2z}}{\omega'_1+\omega'_2-i\lambda}  \\
	& \times \left(\frac{1}{\omega_b + \omega'_1-i\lambda} \; + \; \{\textbf{k}_1 \leftrightarrow \textbf{k}_2\}\right)e^{i(\omega'_1+\omega'_2)t} \, \nonumber, \label{eq:c_22}
	\end{align}
	where
	\begin{align}
	\omega'_{i} &= \Omega_s(k_i)-k_i \, v \, \cos(\theta_i) \, , \quad i = 1,2 \, .
	\end{align}
	The last necessary coefficient is found two have two main contributions $c_1^{(3)}(t) = c_{1,A}^{(3)} + c_{1,B}^{(3)}$ , with
	\begin{align}
	c_{1,A}^{(3)}(t) &= \frac{id^3\phi_{k}^*e^{-kz}}{\omega_b+\omega'_{k}-i\lambda}e^{i\omega_bt+i\omega'_{k}t} \nonumber \\
	&\times \int d^2k' \; \frac{\left(\boldsymbol{\eta} \cdot \vec{k}'\right)\left(\vec{k} \cdot \vec{k}' \right)^*\left|\phi_{k'}\right|^2e^{-2k'z}}{\omega'_k+\omega'_{k'}-i\lambda}\nonumber \\
	&\times \left(\frac{1}{\omega_b + \omega'_{k}-i\lambda} \; + \; \{\textbf{k} \leftrightarrow \textbf{k}'\}\right)\, , \label{eq:c_13_first_term} \\
	c_{1,B}^{(3)} (t) &= c_1^{(1)}(t)\left(c_0^{(2)}(t) +\frac{c_0^{(2)}(t)}{t}\;\frac{i}{\omega_b+\omega_k'-i\lambda}\right) \, .\label{eq:c_13_second_term}
	\end{align}
	
	The expectation value of quantum frictional force can now be evaluated from the expressions for the force operator \eqref{eq:the force operator} and the state of the system at time $t$ \eqref{eq: Atom+Field States}, with the help of the perturbative coefficients
	introduced above. The result is,
	\begin{align}
	\braket{F_x} &= \braket{\Psi(t)| \hat{F}_x |\Psi(t)}\\
	&\approx 2 {\rm Re} \biggl\{\sum_{\eta} \int d\textbf{k}\; \braket{g,0| \hat{F}_x| \boldsymbol{\eta}, \textbf{k}} \nonumber\\
	&\times \left[c_1^{(1)}(t) +c_0^{(2)*}(t)\;c_1^{(1)}(t)+c_1^{(3)}(t)\right]  \nonumber\\
	&+\frac{1}{2}\sum_{\eta}\int d\textbf{k}_1d\textbf{k}_2\; \braket{\boldsymbol{\eta}, \textbf{k}| \hat{F}_x |g, \textbf{k}_1\textbf{k}_2}c_1^{(1)*}(t)\;c_2^{(2)}(t)\biggr\}. \label{eq:F_up_to_the_4th_order}
	\end{align}
	The previous expression is valid up to order $d^4$. From this expression
	we can extract second and fourth order contributions, which will be calculated in the following sections.
	%
	%
	%
	
	\section{Second order frictional force} \label{sec:2ndorder}
	
	We can identify from Eq.~\eqref{eq:F_up_to_the_4th_order} the contribution to order $d^2$, which is given by
	\begin{align}
	\braket{F_x}^{(2)} &= 2{\rm Re}\sum_{\eta}\int d^2 k \; \braket{g,0| \hat{F}_x |\boldsymbol{\eta}, \textbf{k}} \, c_1^{(1)} (t), \label{eq:seonder_order_friction_beginning_step}
	\end{align}
	By explicitly writing out the coefficient $c_1^{(1)}$ \eqref{eq:c_11} and the force operator for the HD model \eqref{eq:the force operator}, we can take the limit $\lambda \rightarrow 0$ and obtain the following expression in polar coordinates:
	\begin{align}
	\braket{F_x}^{(2)} &= d^2 \int_{0}^{\infty} dk \; \bigg[k^3 e^{-2kz} \frac{\omega_p^4}{ \Omega_s(k)\left(\omega_p^2+2\Omega_s^2(k)\right)}  \nonumber\\
	& \times \int_{0}^{2\pi}d\theta\;\cos(\theta) \delta\left(\omega_b+\Omega_s(k)-k\cos(\theta)v\right)\bigg]. \label{eq:second_order_with_delta_function}
	\end{align}
	The integral over $\theta$ can be solved by noting, first, that the $\delta$ function will enforce
	that $\cos(\theta) = \frac{\omega_b+\Omega_s(k)}{kv}$, which is positive. So we can restrict the integral over $\theta$
	to an interval in which $\cos(\theta)>0$. Moreover, since the integrand is even as a function of $\theta$,
	we can write the integral as twice the integral between $0$ and $\pi/2$.
	Then the integral over $\theta$ is performed by making the substitution $y =\cos(\theta)$,
	leading to the appearance of a Heaviside step function $\Theta(-\omega_b-\Omega_s+kv)$,
	due to the fact that $\cos(\theta) \le 1$.
	
	Thereby, our second-order force turns into
	\begin{align}
	\braket{F_x}^{(2)} &= 2d^2 \int_{0}^{\infty}dk \; \Theta(-\omega_b-\Omega_s(k)+kv)\nonumber\\
	& \times  \frac{k^2\omega_p^4e^{-2kz}}{\Omega_s(k)\left(\omega_p^2+2\Omega_s^2(k)\right)}\frac{\omega_b+\Omega_s(k)}{v\sqrt{k^2v^2-\left(\omega_b+\Omega_s(k)\right)^2}} \, . \label{eq:Friction_without_DeltaFunct}
	\end{align}
	With the dispersion relation \eqref{eq:dispersion_relation}, we can perform a change of variables
	in the remaining integral by replacing $k$ by $\Omega_s$,
	\begin{align}
	k = \frac{2\Omega_s^2-\omega_p^2}{2\beta\Omega_s}, \;\; dk = \frac{2\Omega_s^2+\omega_p^2}{2\beta\Omega_s^2} d\Omega_s \label{eq:express_Omega_s_by_k}.
	\end{align}
	By also defining the following dimensionless variables,
	\begin{align}
	u \equiv  \frac{v}{\beta} \, , \quad
	\tilde{\omega} \equiv  \frac{\omega_p}{\omega_b} \, , \quad
	\tilde{z} \equiv  \frac{z \omega_p}{\beta} \, \quad
	w \equiv  \frac{\Omega_s}{\omega_p}
	\label{eq:replace_k_by_Omega}
	\end{align}
	we can rewrite the second-order expectation value of the force operator as
	\begin{align}
	\braket{F_x}^{(2)} &= \frac{d^2\omega_p^4}{2u\beta^4} \int_{\frac{1}{\sqrt{2}}}^{\infty} dw \; \Theta\left[-\omega_b-\Omega_s(w)+k(w) \; v\right] \nonumber\\
	& \times \frac{e^{-\left(2w-\frac{1}{w}\right)\tilde{z}} \; \left(2w^2-1\right)^2\left(1+\tilde{\omega}w\right)}{w^4\sqrt{\left(2w^2-1\right)^2u^2\tilde{\omega}^2-4w^2\left(1 \;+ \; \tilde{\omega}w\right)^2}}. \label{eq:friction_after_substitution}
	\end{align}
	
	The Heaviside function in Eq.~\eqref{eq:friction_after_substitution} modifies the limits of the integration over $w$.
	In App.~\ref{sec:appA} we show an analysis of the restrictions imposed by $\Theta\left(-\omega_b-\Omega_s+kv\right)$,
	and find that they are different in the cases $u \leq 1$ and $u > 1$. We will start by considering the case of small velocities.
	
	For velocities of the particle smaller than the sound speed inside the material ($v \leq \beta$ and hence $u\leq 1$), we find
	that the conditions required by the Heaviside function can never be fulfilled, and hence the integral vanishes.
	This results in an exactly vanishing frictional force for small velocities:
	\begin{equation}
	\braket{F_x}^{(2)} = 0 \quad (u\leq 1).
	\end{equation}
	To this order in perturbation theory, only one (virtual) photon can be excited out of the (dressed) vacuum by the interaction with
	the moving atom. For small enough velocities of the atom, the energy provided by the external source that keeps the particle moving at constant speed
	is not enough to excite the material, and hence
	no friction is produced. A similar threshold was encountered in the case of graphene \cite{farias_graphene}, where for relative velocities
	between the plates smaller than the Fermi velocity of graphene, no fermion could be excited in the material and hence the frictional force
	vanished at the lowest perturbative order.
	
	In the perspective of the HD model, we can attempt a simple pictorial interpretation of this result.
	If the atom is moving, the electrons in the plate will rearrange so that its reflection (formed by image charges) can catch up with the motion of atom. How fast the electrons can rearrange and propagate in the material is limited by the sound speed. For the case of $v \le \beta$, the atom's reflection will move as fast as the atom. Thus, the force between the dipole and its reflection will point always on
	the perpendicular axis, and the frictional force will vanish. In the view of energy conservation, the delta function in Eq.~\eqref{eq:second_order_with_delta_function} gives a resonant condition. With the dispersion relation \eqref{eq:dispersion_relation}, the resonance requires
	\begin{align}
	\omega_b + \frac{1}{2} \left(\sqrt{2\omega_p^2+\beta^2k^2} -\beta k \right) + \left(\beta- v\cos\theta \right) k  = 0.
	\end{align}
	The first two terms are positive, so the third term must be negative, which implies that $v > \beta$ is
	a necessary condition to achieve resonance.

	If the atom velocity exceeds the sound speed ($u > 1$), the restrictions imposed by the Heaviside function
	can be fulfilled in the integration region, and we find a non-vanishing second-order frictional force.
	
	We can normalize such a frictional force by defining ${\braket{f_x}^{(2)} \equiv \braket{F_x}^{(2)} / F_{\rm CP}}$, where $F_{\rm CP} = -\frac{3\alpha}{2\pi z^5} c$ is the static Casimir-Polder force between a perfect conductor and an atom with polarizability $\alpha = 2d^2/\omega_b$ \cite{Casimir1948RetardantVdWs}.
	Then, the normalized second-order frictional force becomes
	\begin{align}
	\braket{f_x}^{(2)} &=\frac{\pi}{6} \frac{\tilde{z}^5}{u \, \tilde{\omega}} \frac{\beta}{c} \int_{w_0}^{\infty} dw \;e^{-\left(2w-\frac{1}{w}\right)\tilde{z}} \label{eq:normalized_second_order_force} \\
	&\;\;\; \times \frac{\left(2w^2-1\right)^2\left(1+\tilde{\omega}w\right)}{w^4\sqrt{\left(2w^2-1\right)^2u^2\tilde{\omega}^2-4w^2\left(1 \;+ \; \tilde{\omega}w\right)^2}} \, , \nonumber
	\end{align}
	where the lower limit $w_0 = [ 1 + \sqrt{1+2\tilde{\omega}^2u\left(u-1\right)}]/{2\tilde{\omega}(u-1)}$
	is a consequence of the Heaviside function, as shown in the Appendix.
	
	The integral in Eq.~\eqref{eq:normalized_second_order_force} is convergent and can be solved numerically. To do so, we shall consider the parameters for the model shown in Table \ref{tab: magnitude}, where we have considered typical values for metals and experimental setups
	that are well within the limits in which we are working, namely, non-retarded and near-field limits.

	\begin{table}[h]
		\centering
		\begin{tabular}{|c|c|}
			\hline
			Model parameter & Range \\
			\hline
			Plasma Frequency $(\omega_p)$ & $10^{15}-10^{16} \text{ s}^{-1}$ \\
			Sound Speed $(\beta)$ & $10^6-10^7 \text{ ms}^{-1}$ \\
			Gap Distance $(z)$    & $10-100 \text{ nm}$  \\
			Bohr Transition Frequency$(\omega_b)$   & $10^{15}-10^{16} \text{ s}^{-1}$ \\
			\hline
			Dimensionless parameter & Range \\
			\hline
			$u\;(v/\beta)$ & $1-20$ \\
			$\tilde{\omega} \;(\omega_p/\omega_b)$ & $0.5-10$ \\
			$\tilde{z} \; (z \omega_p/\beta)$ & $10 - 1000$ \\
			\hline
		\end{tabular}
		\caption{Range of magnitudes considered for the different parameters of the model.}
		\label{tab: magnitude}
	\end{table}

	\begin{figure}
		\includegraphics[width = \columnwidth]{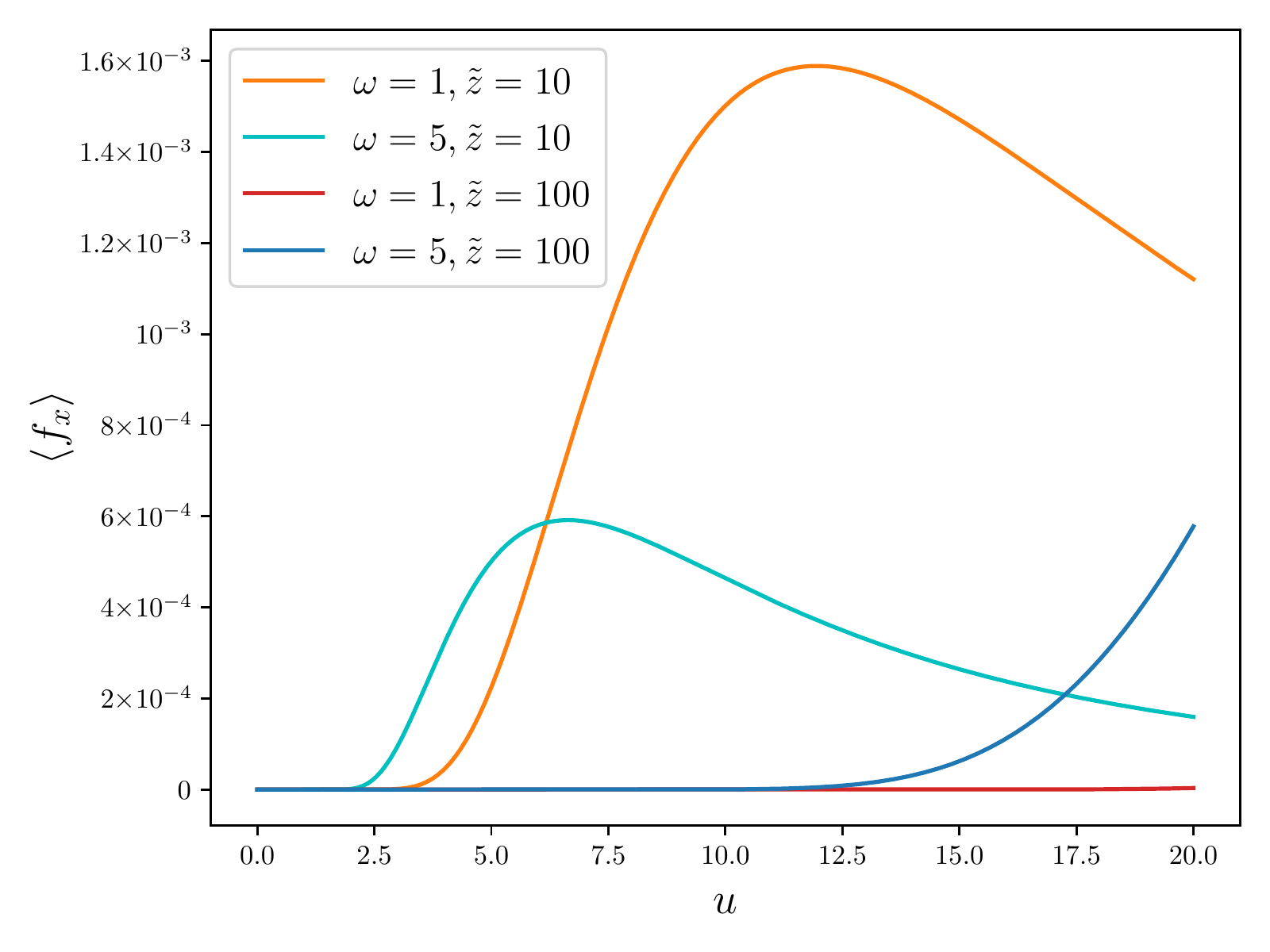}
		\caption{\label{fig:plot} (Color online): Frictional force as a function of the relative velocity between the atom and the plate. The force is normalized by the static Casimir-Polder
		force between a perfectly conducting plate and an atom, and the velocity is normalized as $u = v / \beta$, where $\beta$ is the sound speed in the material.
		We have considered parameters such that $\omega = \omega_p / \omega_b = 1, 5$ and $\tilde{z} = z \omega_p / \beta = 10, 100$. We have taken the sound speed as $\beta = 10^6 \text{ m/s}$. }
	\end{figure}
	
	In Fig. \ref{fig:plot} we show the normalized frictional force as a function of the dimensionless velocity of the atom. The expected behaviour of growing with the velocity and decreasing
	with the gap distance is observed. We have already shown analytically the presence of a threshold that results in a non-vanishing force only for $v > \beta$. However, from the plots shown in the
	figure we can see that the force can be exponentially vanishing even for larger velocities. This effective threshold grows with the gap distance, and decreases with the plasma frequency $\omega_p$.
	As mentioned early, the sound speed has magnitude around $10^6 \text{ m/s}$ which is approximately $1\%$ of the light speed in vacuum.
	To find out whether quantum friction can also be present at lower speeds,
	in the next Section we will have a look at the fourth order force in perturbation theory.

	Finally, we would like briefly discuss the non-dispersive limit. In the limit $\beta \to 0$,
	the dispersion relation \eqref{eq:dispersion_relation} results in a momentum-independent surface-plasmon frequency
	$\Omega_s = \omega_p/\sqrt{2}$. We can recalculate the second-order force accordingly, beginning with Eq.~\eqref{eq:Friction_without_DeltaFunct}, and arrive at the result
		\begin{align}
	\braket{f_x}_{\beta=0}^{(2)} = \frac{\sqrt{2} \pi z^5 \omega_p \omega_b k_0^3}{3vc} \left[K_2(2zk_0) - \left(2zk_0\right)^{-1} K_1(2zk_0)\right], \label{eq:zero_soundspeed_normalized_2nd_force_analytical_result}
	\end{align}
	where $K_n(z)$ are the modified Bessel functions of the second kind and $v k_0 = \omega_b + \omega_p/\sqrt{2}$.
	Hence, within the HD model one obtains a finite frictional force even in the non-dispersive limit.

	
	\section{Fourth order frictional force} \label{sec:4thorder}

	From Eq.~\eqref{eq:F_up_to_the_4th_order}, we can extract the terms of order $d^4$, and split the force into two contributions: one arising from processes
	involving a transition through vacuum $\braket{F_x}_{0}^{(4)}$, and one from processes involving the creation of two photons $\braket{F_x}_{2}^{(4)}$ \cite{intravaia_acceleration}: $\braket{F_x}^{(4)} = \braket{F_x}_{0}^{(4)} + \braket{F_x}_{2}^{(4)}$, with
    \begin{widetext}
	\begin{align}
	\braket{F_x}_{0}^{(4)} &= 2 \text{Re}\sum_{\eta} \int d\textbf{k} \; \braket{g,0|\hat{F}_x|\boldsymbol{\eta},\textbf{k}}
	\left(c_0^{(2)*}(t)\;c_1^{(1)}(t)+ c_{1,B}^{(3)}\right), \label{eq:fourth_order_friction_vaccum}  \\
	\braket{F_x}_{2}^{(4)} &=  2 \text{Re} \bigg\{\sum_{\eta} \int d\textbf{k} \; \braket{g,0|\hat{F}_x|\boldsymbol{\eta},\textbf{k}}c_{1,A}^{(3)}
	+ \frac{1}{2}\sum_{\eta}\int d\textbf{k}_1d\textbf{k}_2 \braket{\boldsymbol{\eta},\textbf{k}|\hat{F}_x|g,\textbf{k}_1\textbf{k}_2}c_1^{(1)*}(t) \; c_2^{(2)}(t)\bigg\} . \label{eq:fourth_order_friction_two_photons}
	\end{align}
	\end{widetext}
	The fourth order frictional force can be cast in the following form, as was shown in Ref.~\cite{intravaia_acceleration}:
	\begin{align}
	\braket{F_x}^{(4)} &\approx \; -\gamma_gt \braket{F_x}^{(2)} - \frac{\partial }{\partial v}\gamma_g  \delta\omega_g + \braket{F_x}_{2}^{(4)}, \label{eq:fourth_order_general_expression}
	\end{align}
	where
	\begin{align}
	\braket{F_x}_{2}^{(4)} =& - \pi d^4\sum_{\eta} \int d^2k_1\;d^2k_2 \; \left|\vec{k}_1 \cdot \vec{k}_2\right|^2 \left|\phi_{k_1}\right|^2\left|\phi_{k_2}\right|^2 \nonumber \\
	&
	\times e^{-2k_1z-2k_2z} \delta\left(\omega'_{1}+\omega'_{2}\right) \; \left[k_1\cos(\theta_1)+k_2\cos(\theta_2)\right]\nonumber \\
	&
	\times \left[\frac{2\omega_b + \omega'_{1} + \omega'_{2} }{\left(\omega_b+\omega'_{1}\right)\left(\omega_b+\omega'_{2}\right)}\right]^2, 
	\end{align}
	and $\gamma_g$ is defined as in Eq.~\eqref{eq:gamma}. This general expression is valid for any model describing the material.
	
	In the case of HD model, when the atom's velocity is slower than the sound speed ($v < \beta$), the first term in Eq.~\eqref{eq:fourth_order_general_expression} vanishes
	due to the fact that the second-order force does. The Dirac $\delta$-function in the definition of $\gamma_g$ Eq.~\eqref{eq:gamma}
	enforces the second term to vanish as well (in the same way it does so for the second order force, as was shown in Sec. \ref{sec:2ndorder}).
	
	The only term that might survive in the case $v< \beta$ is $\braket{F_x}_{2}^{(4)}$, which  contains a different $\delta$-function, namely,
	$\delta(\omega'_{1}+\omega'_{2})$. This $\delta$-function enforces processes in which two photons are created, with momenta
	$\textbf{k}_1$ and $\textbf{k}_2$. Such processes have been shown in other materials to occur at low speeds, since there are more versatile in views of energy and momentum conservation, thus leading to dissipation at low velocities.
	
	However in this case  we find that the peak of the $\delta$-function, $\omega'_1(k_1) + \omega'_2(k_2)=0$, is located outside the
	integration region.
	By explicitly using the dispersion relation \eqref{eq:dispersion_relation},
	we can write the condition imposed by the $\delta$-function as
	\begin{align}
	\sum_{i=1,2}\Omega_s(k_i) \left(1 - \frac{2-1/w_i^2}{2}u\cos(\theta_i)\right) &= 0. \label{eq:delta_function_via_two_photons}
	\end{align}
	From the dispersion relation, we know that the frequency of the surface mode $\Omega_s(k_i)$ must satisfy
	$\Omega_s(k_i) > \omega_p/\sqrt{2}$. Thus, if the above Eq.~\eqref{eq:delta_function_via_two_photons} vanishes,
	it must be as a consequence of the factors between parentheses.
	However, for a slow-moving atom ($u < 1$) and recalling that $w_i > 1/\sqrt{2}$, we see that
	\begin{align}
	\frac{2-1/w_i^2}{2}u\cos(\theta_i) < 1 \, ,
	\end{align}
	which means that Eq.~\eqref{eq:delta_function_via_two_photons} cannot be satisfied. As a consequence, for $v<\beta$ the fourth perturbative order frictional force in the HD model also vanishes.

	Since the $\delta$-functions are determined by conservation laws, we can conclude that, when going to higher perturbative orders,
	we shall find the same type of $\delta$-functions imposing analogous energy conservation. As a consequence, the frictional force
	will vanish at all orders for velocities smaller than the sound speed $v \leq \beta$.
	
	This result is in contrast to what is obtained in other models for metals as the Drude model. The presence of intrinsic
	dissipation in the material results in a frictional force at any velocity. Even though the second-order frictional force
	is found to be vanishingly small for small velocities in the local models considered in the literature, no threshold is encountered.
	The addition of an intrinsic dissipation in the model is not a condition for the existence of friction for higher velocities
	in the case of non-local models, but its absence does impose a threshold on the velocities, as was observed already in the case of graphene \cite{farias_graphene}.
	
	Here for the HD model we find that this threshold is still present at fourth order in perturbation theory. In contrast to the result
	for models containing intrinsic dissipation, where a cubic dependence with the velocity was found, we find a vanishing force for small velocities even at higher orders in perturbation theory.

\section{Conclusions}\label{sec:conc}

In this work, we have calculated the quantum frictional force acting on an atom that moves at constant speed in front of a
metallic plate. To model the material that forms the plate, we have used the hydrodynamic model, a simple non-local model that
does not include intrinsic damping but accounts for the Coulomb interactions. The use of a relatively simple model has allowed us to present detailed analytical calculations that
helped to show explicitly the conditions in which the quantum frictional force is suppressed or enhanced.

Indeed, we have found that the quantum frictional force vanishes identically for velocities smaller than the sound speed $\beta$. This is not surprising, since in our model we do not include any internal source of damping, so that an arbitrarily small amount of energy that is injected into the system cannot be dissipated if it is not sufficient
to excite the internal degrees of freedom of the material. An analogous result was found in the case of graphene \cite{farias_graphene}
for the force up to second order in the coupling constant. However, in our case we have explicitly shown that this result remains true
for both the second and the fourth order force. Moreover, we were able to explicitly identify the mathematical structure which gives
rise to such threshold, which allows us to infer that this result will indeed hold true for any order in perturbation theory.

Physically, the origin of this threshold in the HD model can be interpreted as a consequence of the image charges that form in the material due to the changes in the boundary conditions imposed by the
presence of the atom. Since the sound speed $\beta$ determines the velocity with which the electrons in the material can rearrange,
the reflection can always catch up with the atom when it moves a slower speeds, resulting in a static Casimir-Polder force pointing
vertically, which is generated by the interaction of the two fluctuating dipoles (the atom's and its image's).
It is consistent with this picture that when the threshold velocity vanishes ($\beta \to 0$), i.e., in the non-dispersive limit, the quantum frictional force remains nonzero in the second perturbative order.

For velocities above the threshold we find a non-vanishing force even in the absence of
intrinsic damping. We expect that in clean metals at low temperatures, where impurities and phonon contributions to damping can be neglected, the results for quantum friction we presented will be the leading contribution.

\section{Acknowledgements}
We acknowledge financial support from the National Research Fund Luxembourg under Grants CORE C16/MS/11352881/PARTI, CORE C20/MS/14757511/OpenTop, and ATTRACT A14/MS/7556175/MoMeSys.


	\appendix
	
	\section{Analysis of the limits of the integral in the second order frictional force} \label{sec:appA}
	To solve the integral, we need to discuss how the Heaviside function effects its limits. Firstly, the dispersion relation(Eq.~\eqref{eq:dispersion_relation}) modifies $kv \geq \; \omega_b + \Omega_s$ as
	\begin{align}
	\bigg(\frac{2\Omega_s^2-\omega_p^2}{2\beta\Omega_s}\bigg) v &>\;  \omega_b + \Omega_s \\
	2\left(\beta-v\right)\Omega_s^2 + &2\beta\omega_b\Omega_s + \omega_p^2v <\; 0 \, .
	\end{align}
	By extracting a coefficient $\beta \omega_p^2$, and using the dimensionless variables defined in Eq.~\eqref{eq:replace_k_by_Omega}, we can rewrite this equation as
	\begin{equation} \label{eq:app1}
	2\left(1-u\right)w^2 + 2 \frac{1}{\tilde{\omega}} w + \tilde{\omega}^2u <\; 0 \, .
	\end{equation}

	We can define the left-hand side on Eq.~\eqref{eq:app1}
	as a quadratic polynomial $h(w)$,
	\begin{equation}
	h(w) \equiv\; 2\left(1-u\right)w^2 + 2 \frac{1}{\tilde{\omega}} w + \tilde{\omega}^2u \, .
	\label{eq:boundary_function}
	\end{equation}
	Therefore, we need to find its roots which will determine the interval of $w$ for which the Heaviside function does not vanish.
	They are:
	\begin{align}
	w_{\rm root}
	&= \frac{-1 \pm \sqrt{1+2\tilde{\omega}^2u\left(u-1\right)}}{2\tilde{\omega}(1-u)}. \label{eq:Omegas_roots}
	\end{align}
	For the quadratic function $h(\omega)$, (Eq.~\eqref{eq:boundary_function}), the sign of the quadratic term's coefficient is critical to determine whether $h(w)<0$. When it is not negative or $u \leq 1$, both roots are either negative or complex, and complex roots mean $h(w)$ is always positive. If the coefficient is negative or $u > 1$, there will be one positive and one negative root in Eq.~\eqref{eq:Omegas_roots}. Besides, this case has no complex roots.
	
	We can quickly plot these two cases based on their roots' information. 	
	\begin{figure}[t]
	\includegraphics[width=\columnwidth]{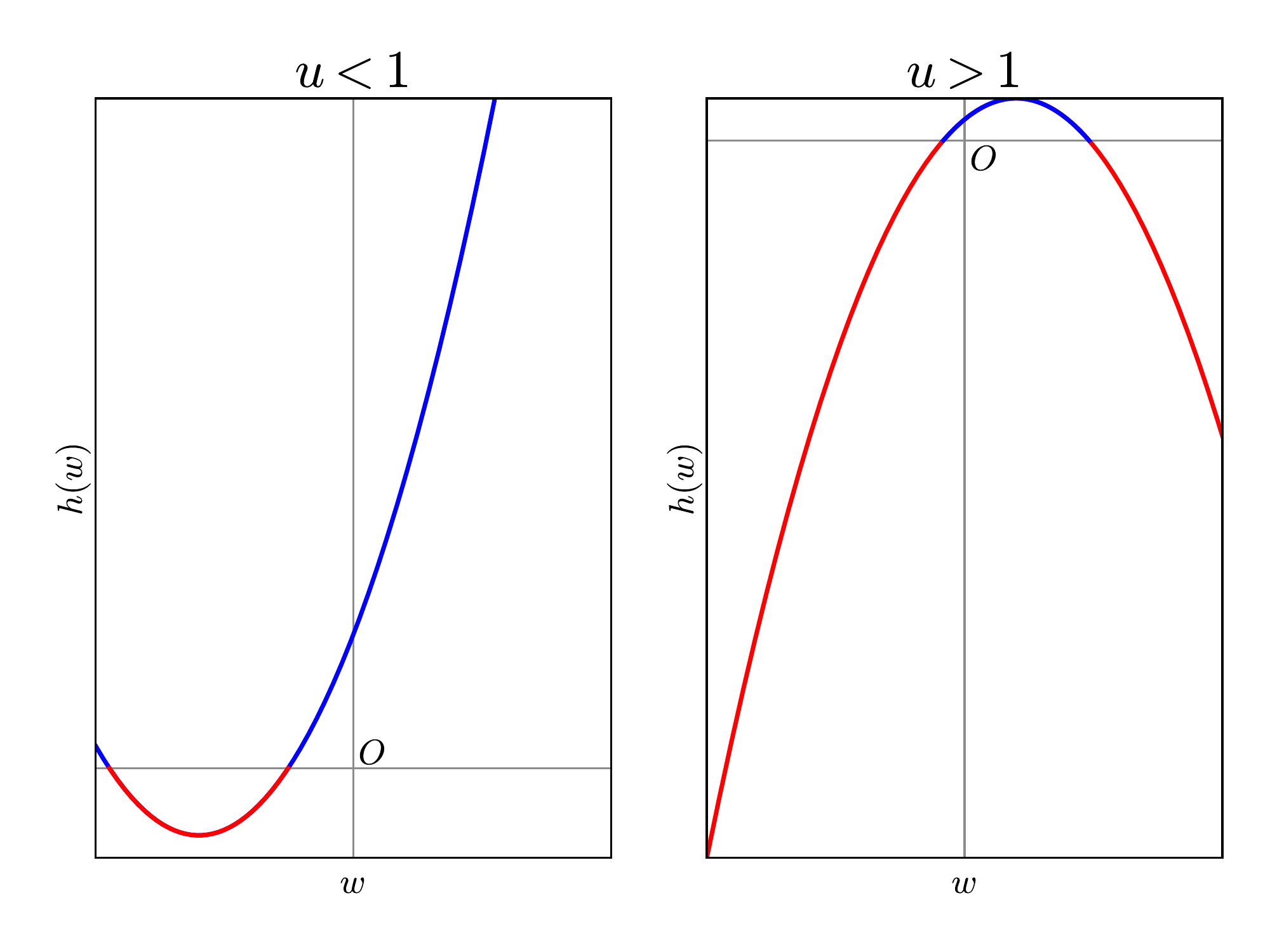}
	\caption{\label{fig:delta_function_analysis} These are possible plots of the quadratic function $h(w)$(Eq.~\eqref{eq:boundary_function}) with $u<1$(left) and $u>1$(right). The red-color region represents for the interval of $w$ giving $h(w)<0$. Alternatively speaking, the Heaviside function does not vanish in these $w$ intervals marked by red. In the case of $u<1$, it is possible that $h(w)$ has no roots, which means $h(w)$ is always positive. However, this case is not what we are interested in. The other case of $u>1$ always has real roots with opposite signs.}
	\end{figure}
	
	The curve marked red is where the Heaviside function does not vanish. However, in the case of $u < 1$ (the left diagram), the range of $w$ corresponding to the red curve requests negative surface-mode frequencies ($\Omega_s<0$). This case gives rise to the integral in Eq.~\eqref{eq:friction_after_substitution} vanishes with a positive range of $\Omega_s$. Thus, the average of the second order force vanishes when $u<1$.

	The other case, $u > 1$, has one positive and one negative root. However, $w$ is the ratio of the surface-mode and plasma frequencies and cannot be negative. So, there is only one interval that would render the Heaviside function as non-vanishing:
	\begin{align}
	w &\in \bigg(\frac{1 + \sqrt{1+2\tilde{\omega}^2u\left(u-1\right)}}{2\tilde{\omega}(u-1)}, \infty \bigg), \; u > 1.
	\end{align}
	To calculate the second order force $\braket{F_x}^{(2)}$, we need to integrate over the interval $(1/\sqrt{2}, \infty)$. This means that we have to compare both lower limits (the one on the integration interval and the one
	imposed by the Heaviside function), and take the lager one or their overlapping region.
	\begin{align}
	w_0 & \equiv\; \frac{1 + \sqrt{1-2\tilde{\omega}^2u+2\tilde{\omega}^2u^2}}{2\tilde{\omega}(u-1)} \\
	&= \frac{1}{\sqrt{2}} \; \frac{1 + \sqrt{1+2\tilde{\omega}^2u\left(u-1\right)}}{\sqrt{2}\tilde{\omega}(u-1)} \\
	& = \frac{1}{\sqrt{2}} \; \sqrt{ \frac{\left(1 + \sqrt{1+2\tilde{\omega}^2u(u -1)}\right)^2}{2\tilde{\omega}^2(u -1)^2}} \\
	& = \frac{1}{\sqrt{2}} \;  \bigg(\frac{1}{\tilde{\omega}^2(u -1)^2}+ \frac{u}{u -1}  \nonumber \\
	& \;\;\;\;\;\;\;\;\;\;\;\;\;\;\;\;\;\;\;\;\;\;\;\; + \frac{ \sqrt{1+2\tilde{\omega}^2u(u -1)}}{\tilde{\omega}^2(u -1)^2}\bigg)^{\frac{1}{2}}\;\;.
	\end{align}
	In the square root of the above formula, there are three terms which are positive, and more importantly, the second term is larger than unity due to $u > 1$. It shows that $w > w_0$ is a subset of $w > 1/\sqrt{2}$. Thus, the lower limit in the expression for $\braket{F_x}^{(2)}$ is replaced by $w_{0}$ from the Heaviside function.

	\bibliography{casimirTI}

\begin{thebibliography}{30}%
\makeatletter
\providecommand \@ifxundefined [1]{%
 \@ifx{#1\undefined}
}%
\providecommand \@ifnum [1]{%
 \ifnum #1\expandafter \@firstoftwo
 \else \expandafter \@secondoftwo
 \fi
}%
\providecommand \@ifx [1]{%
 \ifx #1\expandafter \@firstoftwo
 \else \expandafter \@secondoftwo
 \fi
}%
\providecommand \natexlab [1]{#1}%
\providecommand \enquote  [1]{``#1''}%
\providecommand \bibnamefont  [1]{#1}%
\providecommand \bibfnamefont [1]{#1}%
\providecommand \citenamefont [1]{#1}%
\providecommand \href@noop [0]{\@secondoftwo}%
\providecommand \href [0]{\begingroup \@sanitize@url \@href}%
\providecommand \@href[1]{\@@startlink{#1}\@@href}%
\providecommand \@@href[1]{\endgroup#1\@@endlink}%
\providecommand \@sanitize@url [0]{\catcode `\\12\catcode `\$12\catcode
  `\&12\catcode `\#12\catcode `\^12\catcode `\_12\catcode `\%12\relax}%
\providecommand \@@startlink[1]{}%
\providecommand \@@endlink[0]{}%
\providecommand \url  [0]{\begingroup\@sanitize@url \@url }%
\providecommand \@url [1]{\endgroup\@href {#1}{\urlprefix }}%
\providecommand \urlprefix  [0]{URL }%
\providecommand \Eprint [0]{\href }%
\providecommand \doibase [0]{https://doi.org/}%
\providecommand \selectlanguage [0]{\@gobble}%
\providecommand \bibinfo  [0]{\@secondoftwo}%
\providecommand \bibfield  [0]{\@secondoftwo}%
\providecommand \translation [1]{[#1]}%
\providecommand \BibitemOpen [0]{}%
\providecommand \bibitemStop [0]{}%
\providecommand \bibitemNoStop [0]{.\EOS\space}%
\providecommand \EOS [0]{\spacefactor3000\relax}%
\providecommand \BibitemShut  [1]{\csname bibitem#1\endcsname}%
\let\auto@bib@innerbib\@empty
\bibitem [{\citenamefont {Pendry}(2010)}]{pendry_debate}%
  \BibitemOpen
  \bibfield  {author} {\bibinfo {author} {\bibfnamefont {J.}~\bibnamefont
  {Pendry}},\ }\bibfield  {title} {\bibinfo {title} {Quantum friction--fact or
  fiction?},\ }\href@noop {} {\bibfield  {journal} {\bibinfo  {journal} {New
  Journal of Physics}\ }\textbf {\bibinfo {volume} {12}},\ \bibinfo {pages}
  {033028} (\bibinfo {year} {2010})}\BibitemShut {NoStop}%
\bibitem [{\citenamefont {Philbin}\ and\ \citenamefont
  {Leonhardt}(2009)}]{philbin2009no}%
  \BibitemOpen
  \bibfield  {author} {\bibinfo {author} {\bibfnamefont {T.~G.}\ \bibnamefont
  {Philbin}}\ and\ \bibinfo {author} {\bibfnamefont {U.}~\bibnamefont
  {Leonhardt}},\ }\bibfield  {title} {\bibinfo {title} {No quantum friction
  between uniformly moving plates},\ }\href@noop {} {\bibfield  {journal}
  {\bibinfo  {journal} {New Journal of Physics}\ }\textbf {\bibinfo {volume}
  {11}},\ \bibinfo {pages} {033035} (\bibinfo {year} {2009})}\BibitemShut
  {NoStop}%
\bibitem [{\citenamefont {Casimir}(1948)}]{casimir1948}%
  \BibitemOpen
  \bibfield  {author} {\bibinfo {author} {\bibfnamefont {H.~B.}\ \bibnamefont
  {Casimir}},\ }\bibfield  {title} {\bibinfo {title} {On the attraction between
  two perfectly conducting plates},\ }\href@noop {} {\bibfield  {journal}
  {\bibinfo  {journal} {Front. Phys.}\ }\textbf {\bibinfo {volume} {100}},\
  \bibinfo {pages} {61} (\bibinfo {year} {1948})}\BibitemShut {NoStop}%
\bibitem [{\citenamefont {Milton}(2004)}]{milton2004casimir}%
  \BibitemOpen
  \bibfield  {author} {\bibinfo {author} {\bibfnamefont {K.~A.}\ \bibnamefont
  {Milton}},\ }\bibfield  {title} {\bibinfo {title} {The {Casimir} effect:
  recent controversies and progress},\ }\href@noop {} {\bibfield  {journal}
  {\bibinfo  {journal} {J. of Phys. A}\ }\textbf {\bibinfo {volume} {37}},\
  \bibinfo {pages} {R209} (\bibinfo {year} {2004})}\BibitemShut {NoStop}%
\bibitem [{\citenamefont {Milton}\ \emph {et~al.}(2016)\citenamefont {Milton},
  \citenamefont {H{\o}ye},\ and\ \citenamefont
  {Brevik}}]{milton2016reviewfriction}%
  \BibitemOpen
  \bibfield  {author} {\bibinfo {author} {\bibfnamefont {K.~A.}\ \bibnamefont
  {Milton}}, \bibinfo {author} {\bibfnamefont {J.~S.}\ \bibnamefont
  {H{\o}ye}},\ and\ \bibinfo {author} {\bibfnamefont {I.}~\bibnamefont
  {Brevik}},\ }\bibfield  {title} {\bibinfo {title} {The reality of {Casimir}
  friction},\ }\href@noop {} {\bibfield  {journal} {\bibinfo  {journal}
  {Symmetry}\ }\textbf {\bibinfo {volume} {8}},\ \bibinfo {pages} {29}
  (\bibinfo {year} {2016})}\BibitemShut {NoStop}%
\bibitem [{\citenamefont {Barton}(2010)}]{barton_atom_halfspace}%
  \BibitemOpen
  \bibfield  {author} {\bibinfo {author} {\bibfnamefont {G.}~\bibnamefont
  {Barton}},\ }\bibfield  {title} {\bibinfo {title} {On van der waals friction.
  ii: Between atom and half-space},\ }\href@noop {} {\bibfield  {journal}
  {\bibinfo  {journal} {New Journal of Physics}\ }\textbf {\bibinfo {volume}
  {12}},\ \bibinfo {pages} {113045} (\bibinfo {year} {2010})}\BibitemShut
  {NoStop}%
\bibitem [{\citenamefont {Pendry}(1997)}]{pendry97}%
  \BibitemOpen
  \bibfield  {author} {\bibinfo {author} {\bibfnamefont {J.}~\bibnamefont
  {Pendry}},\ }\bibfield  {title} {\bibinfo {title} {Shearing the
  vacuum-quantum friction},\ }\href@noop {} {\bibfield  {journal} {\bibinfo
  {journal} {Journal of Physics: Condensed Matter}\ }\textbf {\bibinfo {volume}
  {9}},\ \bibinfo {pages} {10301} (\bibinfo {year} {1997})}\BibitemShut
  {NoStop}%
\bibitem [{\citenamefont {Volokitin}\ and\ \citenamefont
  {Persson}(2007)}]{volokitin_persson}%
  \BibitemOpen
  \bibfield  {author} {\bibinfo {author} {\bibfnamefont {A.}~\bibnamefont
  {Volokitin}}\ and\ \bibinfo {author} {\bibfnamefont {B.~N.}\ \bibnamefont
  {Persson}},\ }\bibfield  {title} {\bibinfo {title} {Near-field radiative heat
  transfer and noncontact friction},\ }\href@noop {} {\bibfield  {journal}
  {\bibinfo  {journal} {Reviews of Modern Physics}\ }\textbf {\bibinfo {volume}
  {79}},\ \bibinfo {pages} {1291} (\bibinfo {year} {2007})}\BibitemShut
  {NoStop}%
\bibitem [{\citenamefont {Intravaia}\ \emph {et~al.}(2015)\citenamefont
  {Intravaia}, \citenamefont {Mkrtchian}, \citenamefont {Buhmann},
  \citenamefont {Scheel}, \citenamefont {Dalvit},\ and\ \citenamefont
  {Henkel}}]{intravaia_acceleration}%
  \BibitemOpen
  \bibfield  {author} {\bibinfo {author} {\bibfnamefont {F.}~\bibnamefont
  {Intravaia}}, \bibinfo {author} {\bibfnamefont {V.~E.}\ \bibnamefont
  {Mkrtchian}}, \bibinfo {author} {\bibfnamefont {S.~Y.}\ \bibnamefont
  {Buhmann}}, \bibinfo {author} {\bibfnamefont {S.}~\bibnamefont {Scheel}},
  \bibinfo {author} {\bibfnamefont {D.~A.}\ \bibnamefont {Dalvit}},\ and\
  \bibinfo {author} {\bibfnamefont {C.}~\bibnamefont {Henkel}},\ }\bibfield
  {title} {\bibinfo {title} {Friction forces on atoms after acceleration},\
  }\href@noop {} {\bibfield  {journal} {\bibinfo  {journal} {Journal of
  Physics: Condensed Matter}\ }\textbf {\bibinfo {volume} {27}},\ \bibinfo
  {pages} {214020} (\bibinfo {year} {2015})}\BibitemShut {NoStop}%
\bibitem [{\citenamefont {Far\'{\i}as}\ \emph {et~al.}(2019)\citenamefont
  {Far\'{\i}as}, \citenamefont {Fosco}, \citenamefont {Lombardo},\ and\
  \citenamefont {Mazzitelli}}]{farias2019motion}%
  \BibitemOpen
  \bibfield  {author} {\bibinfo {author} {\bibfnamefont {M.~B.}\ \bibnamefont
  {Far\'{\i}as}}, \bibinfo {author} {\bibfnamefont {C.~D.}\ \bibnamefont
  {Fosco}}, \bibinfo {author} {\bibfnamefont {F.~C.}\ \bibnamefont
  {Lombardo}},\ and\ \bibinfo {author} {\bibfnamefont {F.~D.}\ \bibnamefont
  {Mazzitelli}},\ }\bibfield  {title} {\bibinfo {title} {Motion induced
  radiation and quantum friction for a moving atom},\ }\href@noop {} {\bibfield
   {journal} {\bibinfo  {journal} {Phys. Rev. D}\ }\textbf {\bibinfo {volume}
  {100}},\ \bibinfo {pages} {036013} (\bibinfo {year} {2019})}\BibitemShut
  {NoStop}%
\bibitem [{\citenamefont {Volokitin}\ and\ \citenamefont
  {Persson}(2003)}]{volokitin2003resonant}%
  \BibitemOpen
  \bibfield  {author} {\bibinfo {author} {\bibfnamefont {A.}~\bibnamefont
  {Volokitin}}\ and\ \bibinfo {author} {\bibfnamefont {B.}~\bibnamefont
  {Persson}},\ }\bibfield  {title} {\bibinfo {title} {Resonant photon tunneling
  enhancement of the van der waals friction},\ }\href@noop {} {\bibfield
  {journal} {\bibinfo  {journal} {Physical review letters}\ }\textbf {\bibinfo
  {volume} {91}},\ \bibinfo {pages} {106101} (\bibinfo {year}
  {2003})}\BibitemShut {NoStop}%
\bibitem [{\citenamefont {Klatt}\ \emph {et~al.}(2017)\citenamefont {Klatt},
  \citenamefont {Far{\'\i}as}, \citenamefont {Dalvit},\ and\ \citenamefont
  {Buhmann}}]{klatt_farias}%
  \BibitemOpen
  \bibfield  {author} {\bibinfo {author} {\bibfnamefont {J.}~\bibnamefont
  {Klatt}}, \bibinfo {author} {\bibfnamefont {M.~B.}\ \bibnamefont
  {Far{\'\i}as}}, \bibinfo {author} {\bibfnamefont {D.~A.~R.}\ \bibnamefont
  {Dalvit}},\ and\ \bibinfo {author} {\bibfnamefont {S.}~\bibnamefont
  {Buhmann}},\ }\bibfield  {title} {\bibinfo {title} {Quantum friction in
  arbitrarily directed motion},\ }\href@noop {} {\bibfield  {journal} {\bibinfo
   {journal} {Physical Review A}\ }\textbf {\bibinfo {volume} {95}},\ \bibinfo
  {pages} {052510} (\bibinfo {year} {2017})}\BibitemShut {NoStop}%
\bibitem [{\citenamefont {Dodonov}(2010)}]{review_dyncas}%
  \BibitemOpen
  \bibfield  {author} {\bibinfo {author} {\bibfnamefont {V.}~\bibnamefont
  {Dodonov}},\ }\bibfield  {title} {\bibinfo {title} {Current status of the
  dynamical casimir effect},\ }\href@noop {} {\bibfield  {journal} {\bibinfo
  {journal} {Physica Scripta}\ }\textbf {\bibinfo {volume} {82}},\ \bibinfo
  {pages} {038105} (\bibinfo {year} {2010})}\BibitemShut {NoStop}%
\bibitem [{\citenamefont {Far{\'\i}as}\ \emph {et~al.}(2015)\citenamefont
  {Far{\'\i}as}, \citenamefont {Fosco}, \citenamefont {Lombardo}, \citenamefont
  {Mazzitelli},\ and\ \citenamefont {L{\'o}pez}}]{farias_friction}%
  \BibitemOpen
  \bibfield  {author} {\bibinfo {author} {\bibfnamefont {M.~B.}\ \bibnamefont
  {Far{\'\i}as}}, \bibinfo {author} {\bibfnamefont {C.~D.}\ \bibnamefont
  {Fosco}}, \bibinfo {author} {\bibfnamefont {F.~C.}\ \bibnamefont {Lombardo}},
  \bibinfo {author} {\bibfnamefont {F.~D.}\ \bibnamefont {Mazzitelli}},\ and\
  \bibinfo {author} {\bibfnamefont {A.~E.~R.}\ \bibnamefont {L{\'o}pez}},\
  }\bibfield  {title} {\bibinfo {title} {Functional approach to quantum
  friction: Effective action and dissipative force},\ }\href@noop {} {\bibfield
   {journal} {\bibinfo  {journal} {Phys. Rev. D}\ }\textbf {\bibinfo {volume}
  {91}},\ \bibinfo {pages} {105020} (\bibinfo {year} {2015})}\BibitemShut
  {NoStop}%
\bibitem [{\citenamefont {Klatt}\ \emph {et~al.}(2016)\citenamefont {Klatt},
  \citenamefont {Bennett},\ and\ \citenamefont {Buhmann}}]{klatt2016}%
  \BibitemOpen
  \bibfield  {author} {\bibinfo {author} {\bibfnamefont {J.}~\bibnamefont
  {Klatt}}, \bibinfo {author} {\bibfnamefont {R.}~\bibnamefont {Bennett}},\
  and\ \bibinfo {author} {\bibfnamefont {S.~Y.}\ \bibnamefont {Buhmann}},\
  }\bibfield  {title} {\bibinfo {title} {Spectroscopic signatures of quantum
  friction},\ }\href@noop {} {\bibfield  {journal} {\bibinfo  {journal}
  {Physical Review A}\ }\textbf {\bibinfo {volume} {94}},\ \bibinfo {pages}
  {063803} (\bibinfo {year} {2016})}\BibitemShut {NoStop}%
\bibitem [{\citenamefont {Intravaia}\ \emph {et~al.}(2014)\citenamefont
  {Intravaia}, \citenamefont {Behunin},\ and\ \citenamefont
  {Dalvit}}]{dalvit_intravaia}%
  \BibitemOpen
  \bibfield  {author} {\bibinfo {author} {\bibfnamefont {F.}~\bibnamefont
  {Intravaia}}, \bibinfo {author} {\bibfnamefont {R.}~\bibnamefont {Behunin}},\
  and\ \bibinfo {author} {\bibfnamefont {D.}~\bibnamefont {Dalvit}},\
  }\bibfield  {title} {\bibinfo {title} {Quantum friction and fluctuation
  theorems},\ }\href@noop {} {\bibfield  {journal} {\bibinfo  {journal}
  {Physical Review A}\ }\textbf {\bibinfo {volume} {89}},\ \bibinfo {pages}
  {050101} (\bibinfo {year} {2014})}\BibitemShut {NoStop}%
\bibitem [{\citenamefont {Reiche}\ \emph {et~al.}(2020)\citenamefont {Reiche},
  \citenamefont {Intravaia}, \citenamefont {Hsiang}, \citenamefont {Busch},\
  and\ \citenamefont {Hu}}]{reiche2020nonequilibrium}%
  \BibitemOpen
  \bibfield  {author} {\bibinfo {author} {\bibfnamefont {D.}~\bibnamefont
  {Reiche}}, \bibinfo {author} {\bibfnamefont {F.}~\bibnamefont {Intravaia}},
  \bibinfo {author} {\bibfnamefont {J.-T.}\ \bibnamefont {Hsiang}}, \bibinfo
  {author} {\bibfnamefont {K.}~\bibnamefont {Busch}},\ and\ \bibinfo {author}
  {\bibfnamefont {B.-L.}\ \bibnamefont {Hu}},\ }\bibfield  {title} {\bibinfo
  {title} {Nonequilibrium thermodynamics of quantum friction},\ }\href@noop {}
  {\bibfield  {journal} {\bibinfo  {journal} {arXiv preprint arXiv:2007.04857}\
  } (\bibinfo {year} {2020})}\BibitemShut {NoStop}%
\bibitem [{\citenamefont {Far{\'\i}as}\ and\ \citenamefont
  {Lombardo}(2016)}]{farias2016}%
  \BibitemOpen
  \bibfield  {author} {\bibinfo {author} {\bibfnamefont {M.~B.}\ \bibnamefont
  {Far{\'\i}as}}\ and\ \bibinfo {author} {\bibfnamefont {F.~C.}\ \bibnamefont
  {Lombardo}},\ }\bibfield  {title} {\bibinfo {title} {Dissipation and
  decoherence effects on a moving particle in front of a dielectric plate},\
  }\href@noop {} {\bibfield  {journal} {\bibinfo  {journal} {Phys. Rev. D}\
  }\textbf {\bibinfo {volume} {93}},\ \bibinfo {pages} {065035} (\bibinfo
  {year} {2016})}\BibitemShut {NoStop}%
\bibitem [{\citenamefont {Viotti}\ \emph {et~al.}(2019)\citenamefont {Viotti},
  \citenamefont {Bel\'en~Far\'{\i}as}, \citenamefont {Villar},\ and\
  \citenamefont {Lombardo}}]{viotti2019thermal}%
  \BibitemOpen
  \bibfield  {author} {\bibinfo {author} {\bibfnamefont {L.}~\bibnamefont
  {Viotti}}, \bibinfo {author} {\bibfnamefont {M.}~\bibnamefont
  {Bel\'en~Far\'{\i}as}}, \bibinfo {author} {\bibfnamefont {P.~I.}\
  \bibnamefont {Villar}},\ and\ \bibinfo {author} {\bibfnamefont {F.~C.}\
  \bibnamefont {Lombardo}},\ }\bibfield  {title} {\bibinfo {title} {Thermal
  corrections to quantum friction and decoherence: A closed-time-path approach
  to atom-surface interaction},\ }\href@noop {} {\bibfield  {journal} {\bibinfo
   {journal} {Phys. Rev. D}\ }\textbf {\bibinfo {volume} {99}},\ \bibinfo
  {pages} {105005} (\bibinfo {year} {2019})}\BibitemShut {NoStop}%
\bibitem [{\citenamefont {Far{\'\i}as}\ \emph {et~al.}(2020)\citenamefont
  {Far{\'\i}as}, \citenamefont {Lombardo}, \citenamefont {Soba}, \citenamefont
  {Villar},\ and\ \citenamefont {Decca}}]{farias2020towards}%
  \BibitemOpen
  \bibfield  {author} {\bibinfo {author} {\bibfnamefont {M.~B.}\ \bibnamefont
  {Far{\'\i}as}}, \bibinfo {author} {\bibfnamefont {F.~C.}\ \bibnamefont
  {Lombardo}}, \bibinfo {author} {\bibfnamefont {A.}~\bibnamefont {Soba}},
  \bibinfo {author} {\bibfnamefont {P.~I.}\ \bibnamefont {Villar}},\ and\
  \bibinfo {author} {\bibfnamefont {R.~S.}\ \bibnamefont {Decca}},\ }\bibfield
  {title} {\bibinfo {title} {Towards detecting traces of non-contact quantum
  friction in the corrections of the accumulated geometric phase},\ }\href@noop
  {} {\bibfield  {journal} {\bibinfo  {journal} {npj Quantum Information}\
  }\textbf {\bibinfo {volume} {6}},\ \bibinfo {pages} {1} (\bibinfo {year}
  {2020})}\BibitemShut {NoStop}%
\bibitem [{\citenamefont {Volokitin}\ and\ \citenamefont
  {Persson}(2011)}]{volokitin2011quantum}%
  \BibitemOpen
  \bibfield  {author} {\bibinfo {author} {\bibfnamefont {A.}~\bibnamefont
  {Volokitin}}\ and\ \bibinfo {author} {\bibfnamefont {B.}~\bibnamefont
  {Persson}},\ }\bibfield  {title} {\bibinfo {title} {Quantum friction},\
  }\href@noop {} {\bibfield  {journal} {\bibinfo  {journal} {Physical review
  letters}\ }\textbf {\bibinfo {volume} {106}},\ \bibinfo {pages} {094502}
  (\bibinfo {year} {2011})}\BibitemShut {NoStop}%
\bibitem [{\citenamefont {Farias}\ \emph {et~al.}(2017)\citenamefont {Farias},
  \citenamefont {Fosco}, \citenamefont {Lombardo},\ and\ \citenamefont
  {Mazzitelli}}]{farias_graphene}%
  \BibitemOpen
  \bibfield  {author} {\bibinfo {author} {\bibfnamefont {M.~B.}\ \bibnamefont
  {Farias}}, \bibinfo {author} {\bibfnamefont {C.~D.}\ \bibnamefont {Fosco}},
  \bibinfo {author} {\bibfnamefont {F.~C.}\ \bibnamefont {Lombardo}},\ and\
  \bibinfo {author} {\bibfnamefont {F.~D.}\ \bibnamefont {Mazzitelli}},\
  }\bibfield  {title} {\bibinfo {title} {Quantum friction between graphene
  sheets},\ }\href@noop {} {\bibfield  {journal} {\bibinfo  {journal} {Phy.
  Rev. D}\ }\textbf {\bibinfo {volume} {95}},\ \bibinfo {pages} {065012}
  (\bibinfo {year} {2017})}\BibitemShut {NoStop}%
\bibitem [{\citenamefont {Shaukat}\ and\ \citenamefont
  {Silveirinha}(2020)}]{shaukat2020drag}%
  \BibitemOpen
  \bibfield  {author} {\bibinfo {author} {\bibfnamefont {M.~I.}\ \bibnamefont
  {Shaukat}}\ and\ \bibinfo {author} {\bibfnamefont {M.~G.}\ \bibnamefont
  {Silveirinha}},\ }\bibfield  {title} {\bibinfo {title} {Drag optical force
  due to a drift-current bias of graphene},\ }in\ \href@noop {} {\emph
  {\bibinfo {booktitle} {Metamaterials XII}}},\ Vol.\ \bibinfo {volume}
  {11344}\ (\bibinfo {organization} {International Society for Optics and
  Photonics},\ \bibinfo {year} {2020})\ p.\ \bibinfo {pages}
  {1134417}\BibitemShut {NoStop}%
\bibitem [{\citenamefont {Farias}\ \emph {et~al.}(2018)\citenamefont {Farias},
  \citenamefont {Kort-Kamp},\ and\ \citenamefont {Dalvit}}]{farias2018quantum}%
  \BibitemOpen
  \bibfield  {author} {\bibinfo {author} {\bibfnamefont {M.~B.}\ \bibnamefont
  {Farias}}, \bibinfo {author} {\bibfnamefont {W.~J.}\ \bibnamefont
  {Kort-Kamp}},\ and\ \bibinfo {author} {\bibfnamefont {D.~A.}\ \bibnamefont
  {Dalvit}},\ }\bibfield  {title} {\bibinfo {title} {Quantum friction in
  two-dimensional topological materials},\ }\href@noop {} {\bibfield  {journal}
  {\bibinfo  {journal} {Phys. Rev. B}\ }\textbf {\bibinfo {volume} {97}},\
  \bibinfo {pages} {161407} (\bibinfo {year} {2018})}\BibitemShut {NoStop}%
\bibitem [{\citenamefont {Wilson}\ \emph {et~al.}(2015)\citenamefont {Wilson},
  \citenamefont {Allocca},\ and\ \citenamefont
  {Galitski}}]{wilson2015repulsive}%
  \BibitemOpen
  \bibfield  {author} {\bibinfo {author} {\bibfnamefont {J.~H.}\ \bibnamefont
  {Wilson}}, \bibinfo {author} {\bibfnamefont {A.~A.}\ \bibnamefont
  {Allocca}},\ and\ \bibinfo {author} {\bibfnamefont {V.}~\bibnamefont
  {Galitski}},\ }\bibfield  {title} {\bibinfo {title} {Repulsive {Casimir}
  force between weyl semimetals},\ }\href@noop {} {\bibfield  {journal}
  {\bibinfo  {journal} {Phys. Rev. B}\ }\textbf {\bibinfo {volume} {91}},\
  \bibinfo {pages} {235115} (\bibinfo {year} {2015})}\BibitemShut {NoStop}%
\bibitem [{\citenamefont {Farias}\ \emph {et~al.}(2020)\citenamefont {Farias},
  \citenamefont {Zyuzin},\ and\ \citenamefont {Schmidt}}]{farias2020}%
  \BibitemOpen
  \bibfield  {author} {\bibinfo {author} {\bibfnamefont {M.~B.}\ \bibnamefont
  {Farias}}, \bibinfo {author} {\bibfnamefont {A.~A.}\ \bibnamefont {Zyuzin}},\
  and\ \bibinfo {author} {\bibfnamefont {T.~L.}\ \bibnamefont {Schmidt}},\
  }\bibfield  {title} {\bibinfo {title} {Casimir force between {{Weyl}}
  semimetals in a chiral medium},\ }\href@noop {} {\bibfield  {journal}
  {\bibinfo  {journal} {Phys. Rev. B}\ }\textbf {\bibinfo {volume} {101}},\
  \bibinfo {pages} {235446} (\bibinfo {year} {2020})}\BibitemShut {NoStop}%
\bibitem [{\citenamefont {Barton}(1979)}]{barton1979hydrodynamicmodel}%
  \BibitemOpen
  \bibfield  {author} {\bibinfo {author} {\bibfnamefont {G.}~\bibnamefont
  {Barton}},\ }\bibfield  {title} {\bibinfo {title} {Some surface effects in
  the hydrodynamic model of metals},\ }\href@noop {} {\bibfield  {journal}
  {\bibinfo  {journal} {Reports on Progress in Physics}\ }\textbf {\bibinfo
  {volume} {42}},\ \bibinfo {pages} {963} (\bibinfo {year} {1979})}\BibitemShut
  {NoStop}%
\bibitem [{\citenamefont {Reiche}\ \emph {et~al.}(2017)\citenamefont {Reiche},
  \citenamefont {Dalvit}, \citenamefont {Busch},\ and\ \citenamefont
  {Intravaia}}]{reiche2017spatial}%
  \BibitemOpen
  \bibfield  {author} {\bibinfo {author} {\bibfnamefont {D.}~\bibnamefont
  {Reiche}}, \bibinfo {author} {\bibfnamefont {D.~A.~R.}\ \bibnamefont
  {Dalvit}}, \bibinfo {author} {\bibfnamefont {K.}~\bibnamefont {Busch}},\ and\
  \bibinfo {author} {\bibfnamefont {F.}~\bibnamefont {Intravaia}},\ }\bibfield
  {title} {\bibinfo {title} {Spatial dispersion in atom-surface quantum
  friction},\ }\href@noop {} {\bibfield  {journal} {\bibinfo  {journal}
  {Physical Review B}\ }\textbf {\bibinfo {volume} {95}},\ \bibinfo {pages}
  {155448} (\bibinfo {year} {2017})}\BibitemShut {NoStop}%
\bibitem [{\citenamefont {Jackson}(1999)}]{book_jackson}%
  \BibitemOpen
  \bibfield  {author} {\bibinfo {author} {\bibfnamefont {J.~D.}\ \bibnamefont
  {Jackson}},\ }\href@noop {} {\emph {\bibinfo {title} {Classical
  Electrodynamics}}}\ (\bibinfo  {publisher} {John Wiley \& Sons},\ \bibinfo
  {year} {1999})\BibitemShut {NoStop}%
\bibitem [{\citenamefont {Casimir}\ and\ \citenamefont
  {Polder}(1948)}]{Casimir1948RetardantVdWs}%
  \BibitemOpen
  \bibfield  {author} {\bibinfo {author} {\bibfnamefont {H.~B.~G.}\
  \bibnamefont {Casimir}}\ and\ \bibinfo {author} {\bibfnamefont
  {D.}~\bibnamefont {Polder}},\ }\bibfield  {title} {\bibinfo {title} {The
  influence of retardation on the london-van der waals forces},\ }\href@noop {}
  {\bibfield  {journal} {\bibinfo  {journal} {Phys. Rev.}\ }\textbf {\bibinfo
  {volume} {73}},\ \bibinfo {pages} {360} (\bibinfo {year} {1948})}\BibitemShut
  {NoStop}%
\end{thebibliography}%

\end{document}